\documentclass[preprint,showpacs,preprintnumbers,amsmath,amssymb]{revtex4}%
\usepackage{graphicx}% Include figure files
\usepackage{dcolumn}% Align table columns on decimal point
\usepackage{bm}% bold math

\newcommand{\be}{\begin{equation}}
\newcommand{\en}{\end{equation}}
\newcommand{\bea}{\begin{eqnarray}}
\newcommand{\ena}{\end{eqnarray}}

\begin{document}

%\preprint{GACG/07/2006}

\title{ Approach to exact solutions of cosmological perturbations: Tachyon field inflation}
\author{Ram\'on Herrera}

\email{ramon.herrera@ucv.cl} \affiliation{ Instituto de
F\'{\i}sica, Pontificia Universidad Cat\'{o}lica de Valpara\'{\i}so,
Casilla 4059, Valpara\'{\i}so, Chile.}

\author{Roberto G. P\'{e}rez}
\email{genaro.perez@gmail.com} \affiliation{ Instituto de
F\'{\i}sica, Pontificia Universidad Cat\'{o}lica de
Valpara\'{\i}so, Casilla 4059, Valpara\'{\i}so, Chile.}

\date{\today}% It is always \today, today,

\begin{abstract}
An inflationary universe scenario in the context of tachyon field
is studied. This study is done from an ansantz for the effective
potential of cosmological perturbations $U(\eta)$.  We describe in great detail  the analytical
solutions of the scalar and tensor perturbations for two different
ansantz in the effective potential of cosmological perturbations; the Easther's model and 
an effective potential similar to power law inflation.
Also we find  from the background equations  that the effective tachyonic potential
 $V(\varphi)$, in both models satisfy 
the properties of a tachyonic potential.
We consider the recent data from the Planck data to
constrain the parameters in our effective potential of
cosmological perturbations.

\end{abstract}

\pacs{98.80.Cq}% PACS, the Physics and Astronomy
\maketitle

\section{Introduction}
It is well known that the inflationary scenario provides an
approach for solving the problems of the standard big bang
model\cite{prob, prob2}. However, the inflationary universe model
can offer a graceful mechanism to explain the large-scale
structure\cite{ls}, and likewise the observed  anisotropy of the
cosmic microwave background (CMB) radiation\cite{cmb,Planck2015}.

On the other hand, implications of string/M-theory to
Friedmann-Roberson-Walker (FRW) models have great attention, and
in particular then we considered a tachyonic field associated with unstable
D-branes\cite{sen1}. However, an  important characteristic of the
tachyon field is that this field might be responsible  of  the
inflationary scenario at early times of the universe\cite{ti,ti2},
and also of the dark energy\cite{tde,tde1}. Nevertheless, the tachyonic
field could at the same time produce an inflationary scenario, and
after generate an epoch dominated by a non-relativistic fluid,
which could add to the dark matter\cite{Gi}. Different effective
tachyonic potentials have been considered in the literature, and
in particular an exponential potential constructed from  the
string theory \cite{Kutasov:2000qp},  modifications of
this\cite{Gerasimov:2000zp} and among others. The characteristic
of the effective tachyonic potential $V(\varphi)$, where $\varphi$
denotes a homogeneous tachyon field are: $\partial
V(\varphi>0)/\partial\varphi<0$,
the potential
$V$ has a global maximum at $\varphi\sim 0$, and $V(|\varphi|\rightarrow\infty)\rightarrow 0$. For a review of
tachyon inflation, see Refs.\cite{Steer:2003yu,15}

In the context of  the  cosmological perturbations, the power
spectrums  of  the scalar perturbations, and tensor perturbations
are originated from quantum fluctuations of the vacuum in an
expanding universe\cite{ml}.  It is well known that these fluctuations have  considerable
amplitudes in the very early universe (Planckian length), and
during the expansion of the universe in the inflationary epoch,
these fluctuations are extended with amplitudes beyond of galactic
scales.  In relation to exact solutions of the cosmological
perturbations and the equations of motion  (background space
times) for a standard scalar field, these can be calculated exactly
in the Sitter inflationary universe, in which a constant 
potential is found from the equations of motion of
background\cite{prob}, and also exact results for the spectrums of
the cosmological perturbations\cite{ml}. Also, in the expansion
power law or power law inflation, we can obtained exact results
from the background equations (exponential potential)\cite{Lu},
and exact calculations for the cosmological perturbations, see
Ref.\cite{st}. Similarly, exact results can be obtained in the
Easther's model, where the effective potential of cosmological
perturbations is $U=0$\cite{Ea}. However, this model has an
spectral index $n_S\equiv 3$ (blue tilt), which is significantly different
from the observational data. Generalizations from these spectrums,
were obtained in Ref.\cite{martin}, however the background
solutions was obtained numerically. Also, some approximations to the
cosmological perturbations are studied  in Refs.\cite{A1, A2}

In the following, we explore an approach  to exact results in a 
tachyonic inflationary model. We propose this possibility in the
context of an ansatz  for the effective potential of cosmological
perturbations, and we describe how this ansatz influences the
dynamics of our models, and the solutions of the cosmological
perturbations.  We follow a similar procedure described in  the
Easther's model, and after we study an effective potential of
cosmological perturbations studied in Ref.\cite{martin}.

The outline of the paper goes as follow: In the sections II y III we 
present a short description of the background equations and the
cosmological perturbations for the tachyon field. In the section
IV we study the  Easther's model. In the section V we give
explicit expressions for the scalar and  tensor spectrums,
considering a specific  ansatz  for the effective potential of
cosmological perturbations. At the end, section VI exhibits our
conclusions. We chose units so that $c=\hbar=1$.

\section{ Background model: tachyon field}
In this section  we give a brief description of an inflationary
model, considering that the tachyonic field  drive inflation.

In the context of the Born-Infeld form, the effective action for
tachyon could be described by \cite{la}
 \be
 \label{a1}
 S=\int\sqrt{-g}\left(\frac{R}{2\kappa}-V(\varphi)\sqrt{1+g^{\mu
 \nu}\partial_{\mu}\varphi\partial_{\nu}\varphi}\right)\,d^4x,
 \en
 where the constant $\kappa$ is denoted by  $\kappa=8\pi G$, $R$ is the 
 4-dimensional Ricci scalar, derived from the metric tensor $g_{\mu\nu}$, and the field $\varphi$ is the
 tachyon scalar field with minimal coupling to gravity. Here, the effective scalar 
 potential of the tachyon scalar field is denoted by $V(\varphi)$.

 By taking the  variation of action of Eq.(\ref{a1}) with respect to the two independent variables
  $g_{\mu\nu}$ and $\varphi $, it leads to two main dynamical equations.
Let us introduce into action (\ref{a1}) the flat
Friedmann-Lemaitre-Robertson-Walker (FLRW) metric, given by
 \be
 ds^2=dt^2-a^2(t)\,dx^{i}dx_{i}
 =a^2(\eta)[d\eta^2-dx^{i}dx_{i}],
 \en
 where $\eta$ is the conformal time defined as $dt=a d\eta$, the quantities  $dx^{i}dx_{i}$
 and $a(t)$ represent the flat three-surface and scale factor, respectively.

The energy-momentum tensor of tachyon field yields
$$
T_{\mu\nu}=\frac{V(\varphi)}{\sqrt{1+\varphi^{,c}\varphi_{,c}}}\,[-g_{\mu\nu}(1+\varphi^{,c}\varphi_{,c})+\varphi_{,\mu}\varphi_{,\nu}],
$$
and admits to be cast into the form of a perfect fluid where
$T^\nu_\mu= \,$diag$(-\rho,p,p,p)$, in which the energy density
$\rho$ and pressure $p$ are given by

\be
\rho=\frac{V(\varphi)}{\sqrt{1-\dot{\varphi}^2}},\,\,\,\,\,\mbox{and
}\,\,\,\;\;\;p=-V(\varphi)\sqrt{1-\dot{\varphi}^2}, \en
respectively. Here, the dots mean derivatives with respect to time
$t$.

If the stress energy of the Universe is dominated for a spatially
homogeneous  tachyon field $\varphi(t)$, then
 the Friedmann equation, is given by
\be \label{fried}
H^2=\frac{\kappa}{3}\frac{V(\varphi)}{\sqrt{1-\dot{\varphi}^2}},
\en and the Raychaudhuri equation in the case of the tachyonic
matter yields
 \be
\label{acce}
\frac{\ddot{a}}{a}=H^2+\dot{H}=\frac{\kappa}{3}\frac{V(\varphi)}{\sqrt{1-
\dot{\varphi}^2}}\left(1-\frac{3}{2}\dot{\varphi}^2\right),
\en where $H$ is the Hubble parameter defined as
$H\equiv\dot{a}/a$.

The  inflationary scenario occurs when $\ddot{a}>0$, then
 from Eq.(\ref{acce}), we get  $\dot{\varphi^2}<2/3$. Note that this
condition for the velocity of the scalar tachyon field is
different from the standard scalar field in which
$\dot{\varphi}^2<V$.

The dynamical equation of motion of the tachyon  field from
Eq.(\ref{a1}) becomes
 \be
 \label{conT}
\ddot{\varphi}+3H\dot{\varphi}(1-\dot{\varphi}^{2})+\frac{1}{V}\frac{dV}{d\varphi}(1-\dot{\varphi}^{2})=0,
\en which is equivalent to the  conservation equation
$\dot{\rho}+3H(\rho+p)=0$.

 Combining Eqs. (\ref{fried}) and (\ref{acce}), we get 
\be \label{evT} \dot{\varphi}^2=-\frac{2}{3}\frac{\dot{H}}{H^2},
\en
 and the
effective potential is given by
 \be
 \label{pot1}
V= \frac{3}{\kappa}H^2\sqrt{1+\frac{2}{3}\frac{\dot{H}}{H^2}}. \en
 Here, we  have used Eqs.(\ref{fried}) and (\ref{evT}).

\section{Perturbations}
In this section we will study the scalar and tensor  perturbations
for our models.  We start with the most general  metric
perturbations

\be ds^2=a^{-2}(\eta)[(1+A)d\eta ^2-2\partial_i B dx^id\eta
-[(1-2R)\delta_{ij}+2\partial_i\partial_j E]dx^idx^j].
\label{me}\en In this metric four amounts are needed  to specify its 
 general nature. These quantities are $A$,
$B$, $R$ and $E$,  are functions of the time and space
coordinates and $\eta$ as before is the conformal time
($\eta=\int\,dt/a(t)$).

It is well known  that the scalar perturbations can be measured by
the intrinsic curvature perturbation $\mathcal{R}$ of the comoving
hypersurfaces, that in the case of the standard scalar field is
given by $ \mathcal{R}=-R-(H/\dot{\phi})\delta \phi $. Here,
$\phi$ is the standard scalar field and $\delta \phi$ represents
the fluctuation of the scalar field.  One important quantity to describe the 
perturbations  is the 
gauge-invariant potential $u$, which  is given by $ u=a[\delta
\phi+(\dot{\phi}/H)R]\equiv -z \mathcal{R}, $ where the new
variable $z$ in the case of the standard scalar field is defined
as \be \label{defz} z\equiv a\dot{\phi}/H. \en Here we note the
quantities $\dot{\phi}$ and the Hubble parameter $H$ (or the scale
factor) are determined by the background field equations.

 On the other hand, the equations of motion of 
the linear scalar perturbation for the standard scalar field can be obtained
from the action given by \cite{ml}  \be \label{act}
S=\frac{1}{2}\int d\eta d^3x \left(u'^2-\delta^{ij}
\partial_i u \partial_j u + \frac{z''}{z}u^2 \right), \en
where a prime denotes differentiation with respect the conformal
time $\eta$.

However,  the action given by Eq.(\ref{act}) is formally
equivalent to the action for a scalar field with  a time-dependent
effective mass $m^2=-z''/z$ in flat space-time, and this
equivalence implicates that one can take into account  the quantum
theory.
%Quantizing the variable $u(\eta,\boldsymbol{x})$, we have

%\be \hat{u}(\eta,\boldsymbol{x})=\int
%\frac{d^3\boldsymbol{k}}{(2\pi)^{3/2}}[u_k(\eta)\hat{a}_
%{\boldsymbol{k}} e^{i\boldsymbol{k}\boldsymbol{x}}+
%u^*_k(\eta)\hat{a}^{\dagger}_{\boldsymbol{k}}
%e^{-i\boldsymbol{k}\boldsymbol{x}}
%],\,\,\,\,\,[\hat{a}_{\boldsymbol{k}},\hat{a}^{\dagger}_{\boldsymbol{l}}]=\delta
%^3
%(\boldsymbol{k}-\boldsymbol{l}),\,\,\hat{a}_{\boldsymbol{k}}|0>=0......\label{con}
%\en

From the action  (\ref{act}), the equation for the Fourier modes
$u_k$, for the standard scalar field can be written as
\begin{equation} \label{muka}
u''_k+\left(k^2-\frac{z''}{z}\right)u_k=0,
\end{equation}
where $k=|\boldsymbol{k}|$ is the modulus of the wavenumber.  In
this form,  the Eq.(\ref{muka}) may be considered as a time
independent Schr$\ddot{o}$dinger equation given by
$u''_k+\left(k^2-U\right)u_k=0$, where the effective potential is
defined as $U=U(\eta)=z''/z$. From the Wronskian condition we have
$u_k^*\,u_k'-u_k\,u_k^{*'}=-i$, that guarantees the commutation
relations from the quantum theory. For a review of the
cosmological perturbations of a standard scalar field, see
Refs.\cite{mukhanov, Ly}

%On the other hand, the asymptotic regimes for the Fourier modes
%are given by \be \label{cond1} u_k \rightarrow \frac{1}{2k}
%e^{-ik\eta},\,\,\,\,\, aH \ll k, \en that represents a plane wave
%(short wavelength)
 %and  \be \label{cond2} u_k
%\propto z ,\,\,\,\,\, aH \gg k, \en the growing mode solution
%(long wavelength), which implies that the curvature perturbation
%prevails constant on the super-horizon scales. In this form the
%asymptotic regimes guarantees that the perturbation behaves like a
%free field well inside the horizon and is fixed or constant at
%super-horizon scales.

In the case of  a tachyonic field,  the calculation of the power
spectrum can be obtained  through a canonical variable, similarly
to the standard scalar field. The variable $z$ to the case of a
tachyon field can be written as \cite{sta, gar, hwa}

\begin{equation}
z=\frac{\sqrt{3}\,a\,\dot{\varphi}}{(1-\dot{\varphi}^2)^{1/2}},\label{zt}
\end{equation}
and the analogous equation to Eq.(\ref{muka}) for the Fourier
modes $u_k$ in the case of a tachyonic fluid is given by
\cite{gar}
\begin{equation}
u''_k+\left(k^2\,(1-\dot{\varphi}^2)-U\right)u_k=0,\label{taq}
\end{equation}
where the term $(1-\dot{\varphi}^2)=c_S^2=\delta p/\delta\rho$,
corresponds to the effective speed of sound, and $U$ as before
corresponds to effective potential $U=U(\eta)=z''/z$. Here, we
note that  for a standard scalar field the effective speed of
sound $c_S=1$.

The asymptotic limits    for the equation of the Fourier modes $u_k$ given by Eq.(\ref{taq}) are: Firstly, we
consider that in the limit when the conformal time
$\eta\rightarrow -\infty$  corresponds to a small scale limit, where the modes are  inside the
horizon. Secondly, we take into account that
in the limit when the conformal time $\eta\rightarrow 0$ represents  to a large
scale, here
the modes to be outside the horizon.

The curvature perturbation  in terms of  a Fourier series can be
written as

\be \mathcal{R}=\int
\frac{d^3\boldsymbol{k}}{(2\pi)^{3/2}}\mathcal{R}_{\boldsymbol{k}}(\eta)
e^{i\boldsymbol{k}\boldsymbol{x}},
\en and the vacuum expectation value is defined as  $
\langle\mathcal{R}_{\boldsymbol{k}}\mathcal{R}_{\boldsymbol{l}}^*\rangle
= \frac{2\pi}{k^3}\mathcal{P}_R
\delta^3(\boldsymbol{k}-\boldsymbol{l}), $ where
$\mathcal{P}_R(k)$ is the power spectrum. In this form, the power
spectrum of the curvature perturbations is given by

\be \mathcal{P}_R(k)=\frac{k^3}{2
\pi^2}\left|\frac{u_k}{z}\right|^2. \label{pp}\en

Now considering the  power spectrum of the curvature perturbations
the scalar index $n_s$ is defined as
\begin{equation}
n_s-1=\frac{d\ln \mathcal{P}_R}{d \ln k}.\label{ns}
\end{equation}

In this way, we note that the evolution  of scalar perturbations
during the inflationary scenario is determined  by the function
$z(\eta)$, but this  function $z(\eta)$ at the same time   is
determined through of the effective potential $U(\eta)=z''/z$. In
this form,  instead of considering a specific function $z(\eta)$,
we can obtain exact solutions from Eqs. (\ref{zt}) and
(\ref{taq}), considering the effective potential $U(\eta)$ itself,
see e.g. \cite{martin}.

On the other hand, the production of tensor perturbation during
the inflationary epoch would generate gravitational waves. The
formalism to study  the quantum  fluctuations in the gravitational
field is similar to the case of the scalar perturbation discussed
above. Here, the  equation for the Fourier modes $v_k$, from the
effective gravitational  action is given by\cite{mukhanov, Ly}
\begin{equation}
v_k''+\left(k^2-\frac{a''}{a}\right)\,v_k=0,\label{g}
\end{equation}
and the power spectrum of the tensor perturbations in terms of the
modes $v_k$, is defined as
\begin{equation}
\mathcal{P}_g=\frac{2k^3}{\pi^2}\left|\frac{v_k}{a}\right|^2,
\end{equation}
where now the tensor spectral index $n_T$, is given by $n_T=d\ln
\mathcal{P}_g/d\ln k$.

For the case of the tachyonic field, the  equation for the Fourier modes
 $v_k$   is
exactly the same as in the standard scalar field  Eq.(\ref{g}),
because in absence of anisotropic stress gravity waves are
decoupled from matter. Also, an important observational quantity
is the tensor to scalar ratio $r$, which is defined as
\begin{equation}
r=\frac{\mathcal{P}_g}{\mathcal{P}_R}.
\end{equation}

The Planck collaboration  published new data for the ratio $r$, and found an upper bound on the
tensor to scalar ratio in which $r_{0.002}<0.11$ (95$\%$ C.L.)\cite{Planck2015},
and this upper bound for the tensor to scalar ratio  is similar to obtained from
the Planck mission  in which $r<0.12$  (95$\%$ C.L.)\cite{Planck2014}.

 In the following we will study different  forms  of the
effective potential of cosmological perturbations $U(\eta)$, in the context of
the Born-Infeld theory.

\section{Easther's model: Tachyon field and exact solutions }
In this section we will describe the Easther's model in the
context of a tachyonic field. In this case, which is the simplest case in
which the function $z(\eta)=$ constant or equivalently the
effective potential $U(\eta)=0$. In particular for the standard
scalar field or originally the Easther's model, the differential
equation for the Hubble parameter from the equation of the Fourier
modes is given by, see Ref.\cite{Ea}
\begin{equation}
\frac{1}{2}+\left(\frac{H_{,\phi}}{H}\right)^2-\frac{H_{,\phi\,\phi}}{H}=0,\label{H1}
\end{equation}
where $_{,\phi}$ denotes differentiation with respect to scalar
field $\phi$. The solution of Eq.(\ref{H1}) is given by $H(\phi)=B
\exp[\phi^2/4+A\phi]$, in which $A$ and $B$ are two integration
constants. In this case the spectrum of density perturbations
becomes $\mathcal{P}_R(k)\propto k^2$, and then the exact spectral
index $n_s=3.$ In this form, for a standard scalar field, the
cosmological quantities are exact, and in particular the spectral
index becomes independent of the scalar field. However,  this
value of the spectral index $n_s$ is disapproved from
observational data.

For the tachyonic field the Eq.(\ref{taq}) can be modified  by
defining a new variable $\Theta_k=u_k/z$, such that
\begin{equation}
\Theta_k''+\frac{2}{z}\,z'\,\Theta_k'+k^2\,(1-\dot{\varphi}^2)\Theta_k=0.\label{Ut}
\end{equation}

Considering that the variable $z$ for the tachyon field is given
by Eq.(\ref{zt}), then the Eq.(\ref{Ut}) can be rewritten as
\begin{equation}
\Theta_k''
+\frac{4}{3}\,\frac{a}{H^2}\,\left(H^4-\frac{4}{9}H_{,\varphi}^2\right)^{-1}\,\left[
\frac{H_{,\varphi\,\varphi}}{H}-\frac{4}{3}\,\left(\frac{H_{,\varphi}}{H}\right)^2-\frac{3}{2}
H^2\right]\Theta_k'+k^2(1-\dot{\varphi}^2)\Theta_k=0.\label{cond2}
\end{equation}

Assuming  that the function $z(\eta)$ is a constant, then from
Eq.(\ref{cond2}) the square brackets is equal to zero in which
\begin{equation}
\frac{H_{,\varphi\,\varphi}}{H}-\frac{4}{3}\,\left(\frac{H_{,\varphi}}{H}\right)^2-\frac{3}{2}H^2=0,\label{cond3}
\end{equation}
and the solution of Eq.(\ref{cond3}) can be written as
\begin{equation}
H(\varphi)=\Im^{-1}\,[\varphi],\label{sol1}
\end{equation}
where the function $\Im^{-1}\,[\varphi]$ corresponds to the
inverse function of $\Im\,[\varphi]$ defined as
$$
\Im\,[\varphi]=\frac{6(4C_1(\varphi+C_2)+9(\varphi+C_2)^{7/3})-9(\varphi+C_2)^{7/3}
\sqrt{4+\frac{9(\varphi+C_2)^{4/3}}{C_1}}
\,_2F_1[\frac{1}{2},\frac{3}{4},\frac{7}{4},-\frac{9(\varphi+C_2)^{4/3}}{4C_1}]}{4C_1
\sqrt{4C_1(\varphi+C_2)^{8/3}+9(\varphi+C_2)^4}},
$$
in which $_2F_1$ is the hypergeometric function. The constants
$C_1$ and $C_2$ are two integration constants. Note the difference
between the  Eqs.(\ref{H1}) and (\ref{cond3}) for the Easther's model, in which the quantity $z$=const..

From Eq.(\ref{pot1}) the effective tachyonic potential is given by
\begin{equation}
V(\varphi)=\frac{3\,H^2}{\kappa}\,\sqrt{1-\frac{4}{9}\left(\frac{H_{,\varphi}}{H^2}\right)^2}=\frac{3\,\Im^{-1}\,[\varphi]^2}{\kappa}\,
\sqrt{1-\frac{4}{9}\left(\frac{\Im^{-1}\,[\varphi]_{,\varphi}}{\Im^{-1}\,[\varphi]^2}\right)^2},\label{pot}
\end{equation}
here we have considered the exact solution $H=H(\varphi)$ given by
Eq.(\ref{sol1}).

For the tachyonic field the inflationary scenario occurs when
\begin{equation}
\dot{\varphi}^2=\frac{4}{9}\,\frac{H_{,\varphi}^2}{H^4}=\frac{4}{9}\left(\frac{\Im^{-1}\,
[\varphi]_{,\varphi}}{\Im^{-1}\,[\varphi]^2}\right)^2<\frac{2}{3}.\label{df}
\end{equation}
Here we have considered Eq.(\ref{evT}) in which
$\dot{\varphi}=-\frac{2}{3}\frac{H_{,\varphi}}{H^2}$.

\begin{figure}[th]
\includegraphics[width=3.5in,angle=0,clip=true]{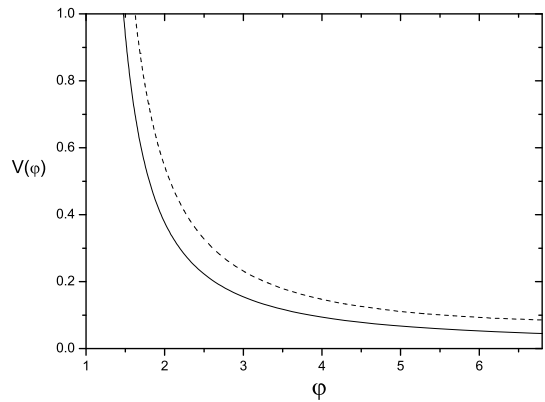}
\caption{ This plot shows the dependence of the effective  tachyonic
 potential $V(\varphi)$ versus the tachyonic field
$\varphi$. Here, we considered two different values of the constant
$C_1$.   The  solid and dashed lines correspond to $C_1=0.1$
and $C_1=0.2$, respectively. In this plot we have used the values
$C_2=1$ and $m_p=1$ .} \label{fig1}
\end{figure}

In Fig\ref{fig1}, we show the evolution of the effective tachyonic
potential  $V(\varphi)$ versus the tachyon field $\varphi$. Here,
we considered two different values of the integration constant
$C_1$. In particular, the solid line corresponds to $C_1=0.1$ and
$C_2=1$,  and the dashed line is for the case in which $C_1=0.2$
and $C_2=1$.

%Also, in the right panel we show the essential condition for that
%the scenario inflationary occurs in which $\dot{\varphi}^2<2/3$
%(the dotted line corresponds to $\dot{\varphi}^2=2/3$). Here the
%tachyon field should begins with a value of $\dot{\varphi}^2<2/3$
%in order to obtain a long period of expansion in which
%$\ddot{a}>0$.
 In order to write down the quantity that relates
$V(\varphi)$,  with the tachyonic field $\varphi$, we consider
Eq.(\ref{pot}), and  we use the values $C_2=1$ and $m_p=1$. From
this plot  we note that the  effective tachyonic  potential
$V(\varphi)$ satisfies the properties: $V_{,\varphi}(\varphi>0)<0$
and $V(\varphi\rightarrow \infty)\rightarrow 0$, that are the
essential conditions for a tachyonic effective potential.

%In particular, from the left panel we observe that for values of
%the tachyon field $\varphi>3$, we obtained a scenario
% inflationary in which $\ddot{a}>0$. In this form,  we find a
% lower bound for the tachyonic field $\varphi$, from the condition $\dot{\phi}^2<2/3$.

\begin{figure}[th]
\includegraphics[width=3.5in,angle=0,clip=true]{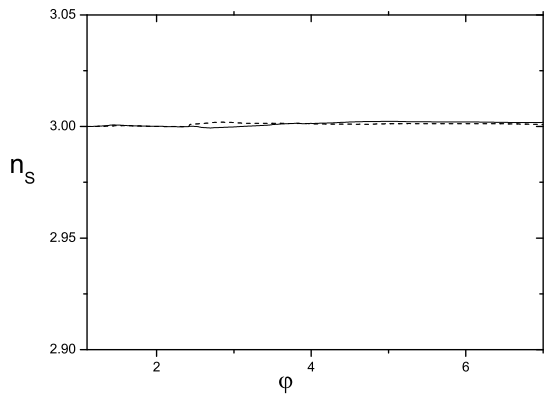}
\caption{ Evolution of the spectral index $n_s$ versus the
tachyonic field $\varphi$. As before, the  solid and dashed  lines
correspond to $C_1=0.1$ and $C_1=0.2$, respectively. In this
plot we have considered the values $C_2=1$ and $m_p=1$}
\label{fig2}
\end{figure}

Although all the quantities from the background field equations
can be found analytically for the case in which the variable $z=$
constant, it  is not possible to determine the
corresponding cosmological perturbations analytically, since we need to use
Eq.(\ref{df}) together  Eq.(\ref{taq}) with $U=0$. In
Fig.\ref{fig2} we show the spectral index $n_s$ versus the
tachyonic field $\varphi$ for the case in which $z$=constant. In
order to write down the spectral index $n_s$, with the tachyonic
field $\varphi$, firstly we consider Eqs.(\ref{taq}) in which
$U=0$ (since the variable $z=$constant) together with the exact
solution given by Eq.(\ref{df}). In this way, we numerically find
the Fourier modes $u_k$, and therefore the power spectrum of the
curvature perturbations for $z=z_0=const.$.  Now, considering
Eq.(\ref{ns}) we numerically obtain the spectral index in terms of
the tachyon field. From Fig.\ref{fig2} we observe that the
spectral index is $n_s\simeq 3$,  but this value is disfavored from  observational
data. Here, we observe  that this results is very similar to
obtained for the standard scalar field, in which $n_S\equiv 3$
(exact result),
  see Ref.\cite{Ea}. This numerical
result for the spectral index in the case of the tachyonic field
is not a surprise, since if we consider that  the
tachyon field slowly changes with the cosmic time, then the
effective speed of sound $c_S^2\sim (1-\dot{\varphi}_0^2)=c^2_
{S_0} =const.$, and the  solution of Eq.(\ref{taq}) for
$z=const.$, is given by $u_k(\eta)=(1/\sqrt{k}) e^{-ic_{S_0}\,k}$,
wherewith $n_s\equiv 3$.

\section{Ansatz: Effective Potential $U(\eta)$}

In this section we will study a specific  ansatz for the effective
potential $U(\eta)$ of cosmological perturbations. Following
Ref.\cite{martin}, a possible   ansatz for the  effective potential is
given by $ U(\eta)= \sum_{m=1}^M \frac{C_m}{|\eta|^m}$,
 where the constants $C_m$'s should be calculated  for different  inflationary model.
For instance  in the  case in which the constants $C_m =
0$ if $m \neq 2$, represents to power-law inflation together with the  inflationary models in the slow-roll
approximation\cite{martin}, and de Sitter
inflation occurs when  $C_2$ is exactly $C_2\equiv 2$.

In this form, considering exact results from  the effective
potential of cosmological perturbations;  power law inflation
 ($a\sim t^p$ with $p>1$) in which
$U(\eta)=\frac{\nu^2-1/4}{\eta^2}$, where  the constant $\nu$ is
given by $\nu=3/2+1/(p-1)$\cite{st}, and for de Sitter inflation
$U(\eta)=\frac{2}{\eta^2}$\cite{mukhanov},
%In order to obtain an exact solution for the variable $z(\eta)$ in the tachyonic field,
we consider the effective potential of cosmological perturbations
can be written as 

\be \label{potenc} U(\eta)=\frac{z''}{z}=
C_0+\frac{C_1}{\eta}+\frac{C_2}{\eta^2}, \en where $C_0$, $C_1$
and $C_2 $ are  free parameters. Here the parameter $C_0$ has dimensions of 
mass-squared, $C_1$ of mass and $C_2$ is dimensionless. 

Considering the change of the variable
$\bar{\eta}=2\sqrt{C_{0}}\eta$, then Eq.(\ref{potenc}) can be
rewritten as
\begin{equation}
\frac{d^{2}z}{d\bar{\eta}^{2}}+\left(-\frac{1}{4}+\frac{\alpha}{\bar{\eta}}+\frac{\frac{1}{4}
-\beta^{2}}{\bar{\eta}^{2}}\right)z(\bar{\eta})=0,\label{W1}
\end{equation}
where the quantities $\alpha$ and $\beta$ are constants, and are
defined as
\begin{equation}\alpha=\frac{-C_1}{2\sqrt{C_0}}, \,\,\,\mbox{and}\;\;\;\,
\beta^2=\frac{1}{4}+C_{2},\label{cons}
\end{equation}
respectively.
 The exact solution of Eq.(\ref{W1})  is
given in terms of Whittaker functions $W_{\pm\alpha,\beta}$
\cite{Wi}, and the solution for the variable $z$ can be written as
\begin{equation}
z(\bar{\eta})=A_1W_{\alpha,\beta}(\bar{\eta})+A_2W_{-\alpha,\beta}(-\bar{\eta}),\label{zet}
\end{equation}
where $A_1$ and $A_2$ are two integration constants. We note that
in general our variable fundamental $z$  is characterized by the
constants $C_0$, $C_1$ and $C_2 $, and two
 integration constants  $A_1$ and $A_2$.

In order to obtain  analytical solutions for the background field equations and the cosmological perturbations
(scalar and tensor perturbations),  we
considered that the tachyonic field slowly changes
with the cosmic time, and the velocity $\dot{\varphi}$ is approximately constant, i.e.,
$\dot{\varphi}^2\sim \dot{\varphi_0}^2=const.<2/3$, then the
effective speed of sound becomes
$c_S^2\sim(1-\dot{\varphi_0}^2)=c_{S_{0}}^2$, where
$c_{S_{0}}^2>1/3$.

Under this approximation, and considering the Eqs.(\ref{zt}) and (\ref{zet}), we
find that the scale factor $a$ is given by
$a(\eta)=\overline{A}_1\,W_{\alpha,\beta}(2\sqrt{C_0}\,\eta)$,
where the constant $\overline{A}_1$ is defined as
$\overline{A}_1=c_{S_0}\,A_1/\sqrt{3(1-c_{S_0}^2)}$, and  from
Eq.(\ref{pot1}) the tachyonic potential in terms of the conformal time $\eta$ can be written as
\begin{equation}
V(\eta)=\frac{3}{\kappa}\,\frac{\mathcal{H}^2}{a^2}\,\sqrt{\frac{1}{3}\left[1+\frac{2\mathcal{H}'}
{\mathcal{H}^2}\right]}=\frac{3}{\kappa\overline{A}_1^2}
\left[\frac{W'_{\alpha,\beta}}{W_{\alpha,\beta}^2}\right]^2\sqrt{\frac{1}{3}\left[1+\frac{2(W_{\alpha,\beta}
W_{\alpha,\beta}^{''}-W_{\alpha,\beta}^{'\,2})}{W_{\alpha,\beta}^{'\,2}}\right]},
\end{equation}
where $\mathcal{H}=Ha=a'/a$, and $W_{\alpha,\beta}$ corresponds to
the Wittaker function $W_{\alpha,\beta}(2\sqrt{C_0}\,\eta)$. Here
we have used that the constant $A_2=0$ (without loss of
generality), and as before a prime denotes differentiation with
respect the conformal time $\eta$. Considering the
$\dot{\varphi}=\varphi'/a\sim\dot{\varphi}_0=(1-c_{S_0})^{1/2}$,
we obtain that $\varphi(\eta)\propto\int W_{\alpha,\beta}d\eta$,
and then we would obtain $V=V(\varphi)$. However, the field
$\varphi(\eta)$ does not have an analytical solution. Instead of
consider this numerical integral, we numerically solve the full
background equations.

\begin{figure}[th]
\includegraphics[width=4.5in,angle=0,clip=true]{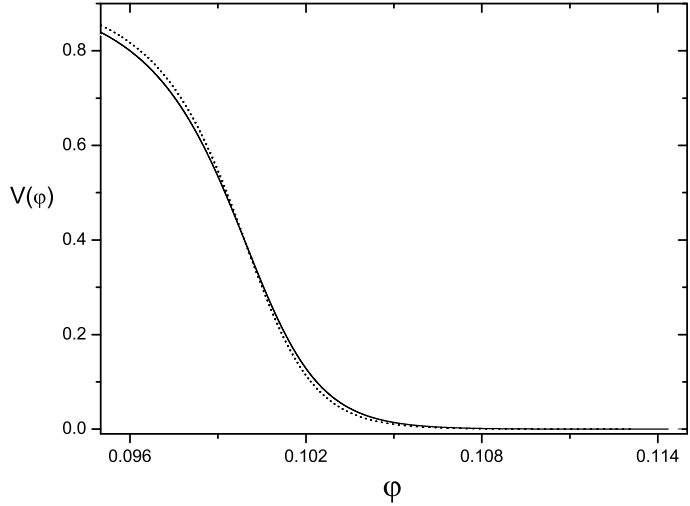}
\caption{ Numerical result: Evolution of the effective potential
$V(\varphi)$ versus the tachyonic field $\varphi$. Here, the doted
and solid lines correspond to $C_1=10^{-5}$ and $C_1=10^{-7}$ ,
respectively. In this plot we have considered the values
$C_0=10^{-8}$, $C_2=2$, $A_1=10^{-6}$ and $m_p=1$.} \label{fig4}
\end{figure}

In Fig\ref{fig4}, we show the evolution of the effective tachyonic
potential  $V(\varphi)$ versus the tachyon field $\varphi$. In
doing this, we considered two different values of the constant
$C_1$. Here, the doted line corresponds to $C_1=10^{-5}$ and solid
line to $C_1=10^{-7}$.  Also, in this plot we have used  the values
$C_0=10^{-8}$, $C_2=2$, $A_1=10^{-6}$ and $m_p=1$.  In order to
write down the effective tachyonic potential $V(\varphi)$ versus
$\varphi$,
 we numerically  utilize
Eqs.(\ref{zt}) and (\ref{potenc}), together with the background
equations given by Eqs.(\ref{fried})-(\ref{conT}). From this plot,
we observe that the tachyonic effective potential $V(\varphi)$
satisfies the properties; $V_{,\varphi}(\varphi>0)<0$ and
$V(\varphi\rightarrow \infty)\rightarrow 0$, that are the
conditions for an effective tachyonic potential. We note that this
tachyonic potential is  analogous  to obtained in  the Easther's
model (see section IV).

Now from Eqs.(\ref{taq}) and (\ref{potenc}), the equation for the
Fourier modes $u_k$, becomes
\begin{equation}
u_{k}''(\eta)+\left(c_{S_0}^2k^2-C_0-\frac{C_1}{\eta}-\frac{C_2}{\eta^2}\right)u_{k}(\eta)=0.\label{u1}
\end{equation}

Redefining the conformal time as
$\tilde{\eta}=2i\sqrt{c_{S_{0}}^2k^2-C_0}\;\eta$, the above
equation results
\begin{equation}
\frac{d^{2}u_{k}}{d\tilde{\eta}^{2}}+\left(-\frac{1}{4}+\frac{\tilde{\alpha}}{\tilde{\eta}}+
\frac{\frac{1}{4}-\beta^{2}}{\tilde{\eta}^{2}}\right)u_{k}=0,\label{u2}
\end{equation}
where now the constant $\tilde{\alpha}$ is defined by
$\tilde{\alpha}=\frac{C_1\,i}{2\sqrt{c_{S_{0}}^2k^2-C_0}}$, and the
constant $\beta$ as before is given  by Eq.(\ref{cons}). From
Eq.(\ref{u2}), we find that the analytical solution  for the
Fourier modes can be written as
\begin{equation}
u_{k}(\tilde{\eta})=B_1W_{\tilde{\alpha},\beta}(\tilde{\eta})+
B_2W_{-\tilde{\alpha},\beta}(-\tilde{\eta}),\label{sol2}
\end{equation}
where $B_1$ and $B_2$ are two integration constants, and whose values 
 are fixed by the initial conditions, and as before
$W_{\pm\tilde{\alpha},\beta}$ correspond to the  Whittaker
functions.

In order to fix the constants $B_1$ and $B_2$, we consider the
 small scale limit in which the conformal time
$\eta\rightarrow -\infty$, then from Eq.(\ref{sol2}) we get
$$
u_{k}(\eta) \rightarrow
\frac{1}{\sqrt{2k_{eff}}}e^{-ik_{eff}\eta},
$$
where $k_{eff}$ corresponds to the effective modulus of
wavenumber, and is given by
$k_{eff}=\sqrt{c_{S_{0}}^2k^2-C_0}\,>0$. Here, we observe that
$k_{eff}$ represents a shift in the effective wavenumber
$c_{S_{0}}k$. Now considering that in the asymptotic limit
$\eta\rightarrow -\infty$ (small scale limit) the Whittaker
function is proportional to
$e^{-\frac{\tilde{\eta}}{2}}\tilde{\eta}^\alpha$, we get

\begin{equation}
B_1=\frac{e^{\frac{\pi C_1}{4k_{eff}}}}{\sqrt{2k_{eff}}},
\,\,\,\mbox{and}\,\,\,\,\,\;B_2=0.\label{c1}
\end{equation}
Here, we have used the Wronskian condition. By using  the
Eqs.(\ref{sol2}) and (\ref{c1}), we find that the analytical
solution for the Fourier modes can be written as

\begin{equation}
u_k(\tilde{\eta})=\frac{e^{\frac{\pi
C_1}{4k_{eff}}}}{\sqrt{2k_{eff}}}W_{\tilde{\alpha},\beta}(\tilde{\eta}).
\end{equation}

On the other hand, we consider the growing modes in the large
scale limit in which the conformal time $\eta\rightarrow 0$ (the
quantum fluctuations are frozen outside the horizon)  the
Whittaker function becomes $
W_{\tilde{\alpha},\beta}(\tilde{\eta})\rightarrow
 \frac{\Gamma(2\beta)}{\Gamma(\frac{1}{2}+\beta-
 \tilde{\alpha})}\tilde{\eta}^{1/2-\beta}e^{-\frac{\tilde{\eta}}{2}},
$ and then  we obtain  that the analytical solution for
$u_k(\eta)$ for large scale can be written as
\begin{equation}
\label{ugranescala}
u_k(\eta)=\frac{e^{\pi C_1/4k_{eff}}}{\sqrt{2k_{eff}}}
\frac{\Gamma(2\beta)}{\Gamma(\frac{1}{2}+\beta-\tilde{\alpha})}
{(2ik_{eff}\eta)}^{1/2-\beta}e^{-ik_{eff}\eta}.
\end{equation}

In this form, considering Eqs.(\ref{pp}) and (\ref{ugranescala}),
we find  that the power spectrum of the curvature perturbations
becomes
\begin{equation}
P_{R}(k)=B \frac{k^3k_{eff}^{-2\beta}}{\Gamma(s)\Gamma^*(s)}e^{\pi
C_1/2k_{eff}},\,\,\,\mbox{where}\,\,\,B=\frac{\Gamma^2(1/2+\beta-\alpha)}{4 \pi^2 A_1^2
C_0^{1/2-\beta}},\,\,\mbox{and}\,\,\,s=\frac{1}{2}+\beta-\tilde{\alpha}.\label{Pr}
\end{equation}
Here, we considered that in the limit   $\bar{\eta}\rightarrow 0$
(large scale limit), the variable $z$  is given by $
z(\bar{\eta})\rightarrow
A_1\frac{\Gamma(2\beta)}{\Gamma(\frac{1}{2}+\beta-\alpha)}\bar{\eta}^{\frac{1}{2}-\beta},
$ and as before the constant $A_2=0$.

From the power spectrum given by Eq.(\ref{Pr}), we find that the spectral scalar index $n_s$, considering
Eq.(\ref{ns}) is given by
\begin{equation}
n_s(k)=4-\frac{c_{s0}^2 k^2}{k_{eff}}\left[2\beta
k_{eff}^{2\beta}+\frac{
C_1}{2k_{eff}^2}\,\left(\pi+i\left[\Psi_0\left(s\right)-\Psi_0\left(s^*\right)\right]\right)\right],\label{ns2}
\end{equation}
where $\Psi_0(s)$ corresponds to the Polygamma function.

In the context of the spectrum of gravity waves for a tachyonic
field, the equation of the Fourier modes $v_k$ is exactly the same as in
the standard scalar field, see Eq.(\ref{g}). In this form, by considering
Eqs. (\ref{zt}) and (\ref{g}), the equation of the Fourier modes
$v_k$ can be written as
\begin{equation}
\frac{d^2v_k}{d\eta_T^2}+\left(-\frac{1}{4}+\frac{\alpha_T}{\eta_T}
-\frac{\frac{1}{4}-\beta^2}{\eta_T^2}\right)\,v_k=0,\label{g2}
\end{equation}
where the constant $\alpha_T=\frac{C_1\,i}{2\sqrt{k^2-C_0}}$,  the constant  $\beta$
 as before is given  by Eq.(\ref{cons}), and the new
conformal time is defined as $\eta_T=2i\sqrt{k^2-C_0}\;\eta$.  In this way,
we obtain that the exact solution for the modes $v_k$ of Eq.(\ref{g2})
becomes
\begin{equation}
v_k(\eta)=D_1\,W_{\alpha_T,\beta}(2i\sqrt{k^2-C_0}\;\eta)+
D_2W_{-\alpha_T,\beta}(-2i\sqrt{k^2-C_0}\;\eta),\label{sol22}
\end{equation}
where $D_1$ and $D_2$ are two integration constants.  In particular  for the Sitter universe,  the scale
factor is $a\propto-\eta^{-1}$, and the equation of the modes
$v_k$ is $v_k''+(k^2-\frac{2}{\eta^2})v_k=0$, that in our case
corresponds to $C_0=0$, $C_1=0$ and $C_2=2$, and  the solution for the modes $v_k$, is given by
$v_k=(1/\eta)[D_1(k\eta \cos[k\eta]-\sin[k\eta])+D_2(k\eta
\sin[k\eta]+\cos[k\eta])]$, see Ref.\cite{mukhanov}.

Following Ref.\cite{mukhanov}, the initial conditions for the modes
$v_k(\eta_i)$ and $v'_k(\eta_i)$ (initial conformal time $\eta_i$) are given by
\begin{equation}
v_k(\eta_i)=\frac{1}{[k^2-C_0-\frac{C_1}{\eta_i}-\frac{C_2}{\eta_i^2}]^{1/4}},\,\,\,\,\mbox{and}\,\,\,
\,v'_k(\eta_i)=i\,\left[k^2-C_0-\frac{C_1}{\eta_i}-\frac{C_2}{\eta_i^2}\right]^{1/4},\label{cc}
\end{equation}
respectively. Here, the initial conditions make sense exclusively
if $[k^2-C_0-\frac{C_1}{\eta_i}-\frac{C_2}{\eta_i^2}]>0$.
Considering the solution of the modes $v_k$, together the initial
conditions given by Eq.(\ref{cc}), we find that the constants
$D_1$ and $D_2$ becomes
$$
D_1=\frac{\alpha_T}{2\sqrt{k_B}}\frac{(2k_B\eta_i)^{-\frac{iC_1}{2k_B}}}{(\alpha_T+ik_B\eta_i)}\,e^{ik_B\eta_i+\frac{\pi
C_1}{4k_B}},\;
\mbox{and}\,\,\,D_2=\frac{1}{2\sqrt{k_B}}\frac{(\alpha_T+2ik_B\eta_i)(-2k_B\eta_i)^{\frac{iC_1}{2k_B}}}{(\alpha_T+ik_B\eta_i)}\,e^{-(ik_B\eta_i+\frac{\pi
C_1}{4k_B})}.
$$
Here, we considered that for $\eta\longrightarrow -\infty$ (small
scale limit), the normalization given by Eq.(\ref{cc}) becomes
$[k^2-C_0-\frac{C_1}{\eta_i}-\frac{C_2}{\eta_i^2}]^{1/4}\simeq
[k^2-C_0]^{1/4}=\sqrt{k_B}>0$, where $k_B=\sqrt{k^2-C_0}$, and
also we assumed that the asymptotic behavior of the Wittaker
function in the small scale limit is
$W_{\alpha_T,\beta}(2ik_B\eta)\rightarrow
(2ik_B\eta)^{\alpha_T}\,e^{-ik_B\eta}$. Also, we note that in
particular for the case in which $C_0=0$, the initial conditions
are given by $v_k(\eta_i)=1/\sqrt{k}$ and
$v'_k(\eta_i)=i\sqrt{k}$, when $\eta\longrightarrow -\infty$ \cite{mukhanov}.

In this form,  from Eqs.(\ref{zet}) and (\ref{sol22}), the power spectrum of the tensor perturbation, is
given by
\begin{equation}
\mathcal{P}_g=\frac{2k^3}{\pi^2}\left|\frac{v_k}{a}\right|^2=\frac{2k^3}{\pi^2\tilde{A}_1^2}\,
\frac{\mid D_1\,W_{\alpha_T,\beta}(2ik_B\;\eta)+
D_2W_{-\alpha_T,\beta}(-2ik_B\;\eta)\mid^2}{\mid W_{\alpha,\beta}(2\sqrt{C_0}\eta
)\mid^2},\,\,\,\label{ex}
\end{equation}
where  the constant $\tilde{A_1}$ is defined as
$\tilde{A}_1=\frac{c_{S_0} \,A_1}{\sqrt{3(1-c_{S_0}^2)}}$.

Considering  the large scale limit in which $\eta\rightarrow 0$,
then the spectrum of the gravitational waves given by
Eq.(\ref{ex}), can be written as

$$
\mathcal{P}_g(k)\simeq\frac{2k^3}{\pi^2\tilde{A_1}^2}\,\mid\Gamma\left(1/2+\beta-\alpha\right)\mid^2\,
\left(\frac{k_B^2}{C_0}\right)^{1/2-\beta}\,\,\,\times\Bigl[\frac{\mid
D_1\mid^2}{\mid\Gamma\left(1/2+\beta-\alpha_T\right)\mid^2}
$$
\begin{equation}
+\frac{\mid
D_2\mid^2}{\mid\Gamma\left(1/2+\beta+\alpha_T\right)\mid^2}+
(-1)^{1/2-\beta}
\,\mbox{Re}\left(\frac{2D_1\,D_2^*}{\Gamma\left(1/2+\beta-\alpha_T\right)
\Gamma^*\left(1/2+\beta+\alpha_T\right)} \right)\Bigr].\label{Pg2}
\end{equation}

Here, we considered that in the large scale limit
($\eta\rightarrow 0$),  the Wittaker function is given by
$W_{\alpha_T,\beta}\rightarrow
[\Gamma(2\beta)/\Gamma(1/2+\beta-\alpha_T)]\,(2ik_B\eta)^{1/2-\beta}\,e^{-ik_B\eta}$.

\begin{figure}[th]
{\hspace{-4
cm}\includegraphics[width=4.5in,angle=0,clip=true]{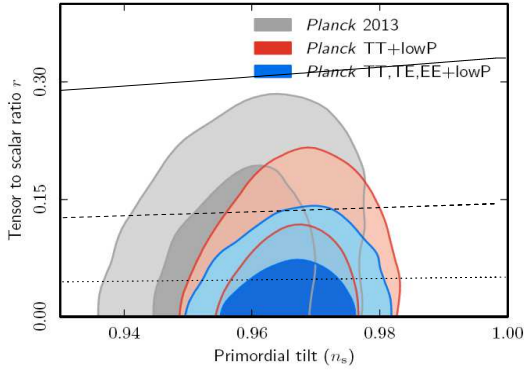}}
{\vspace{-1 cm}\caption{ The plot shows the  tensor to scalar
ratio $r$ versus the scalar spectrum  index $n_s$ in the case of
the tachyonic field $\varphi$, for different values of the speed
of sound $c_{S_0}$. In this plot we show the two-dimensional
marginalized constraints (68$\%$ and 95$\%$ C.L.) on the
parameters $r$ and $n_s$ obtained from the Planck 2015
data\cite{Planck2015}. Here, the solid, dashed, and doted lines
correspond to $c_{S_0}=0.993$, $c_{S_0}=0.997$ and $c_{S_0}=0.999$,
respectively. In this plot we have considered the values
$\eta_i=-400.000$, $C_0=10^{-8}$,$C_1=1$, $C_2=2$ and
$m_p=1$.\label{figg5} }}
\end{figure}

In Fig.\ref{figg5} we show the tensor to scalar ratio $r$ versus
the scalar spectral index $n_s$, for the  effective potential of
cosmological perturbations $U(\eta)$ given by the ansantz  of
Eq.(\ref{potenc}). Here, we show the two-dimensional marginalized
constraints for the tensor to scalar ratio $r$ versus $n_s$, from
the new results of Planck 2015 data\cite{Planck2015}. Combining
Eqs.(\ref{Pr}), (\ref{ns2}) and (\ref{Pg2}), we numerically find
the consistency relation $r=r(n_s)$, and also we consider three
values of the effective
 speed of sound. The solid, dashed, and doted lines correspond to  the speed sound  $c_{S_0}=0.993$,
$c_{S_0}=0.997$ and $c_{S_0}=0.999$, respectively.  Also,
 in this plot we
have used  the values $\eta_i=-400.000$,
$C_0=10^{-8}$,$C_1=1$, $C_2=2$ and $m_p=1$. From the two-dimensional
marginalized constraints  we note that for the value $c_{S_0}>0.993$  the model
is well supported by the new Planck data.  In this form, we observe that for the
case  of de Sitter inflation in which $C_2=2$ (recalled that in de Sitter
$U(\eta)=2/\eta^2$), and for different  values of $C_0$ (with $c_{S_0}^2k^2>C_0$) and $C_1<10^2$, the tachyonic model is very similar to the standard scalar
field,  where $c_S\equiv 1$.  Therefore, it may come as a surprise that exist a similitude in the trajectories
 $n_s-r$ plane between standard field and tachyonic field. However, this  similitude  was noted in Ref.\cite{Steer:2003yu} at lower order
in the slow roll parameters. Also, we observe  that for small
deviations of the value of $C_2=2$ ( for instance, power law
inflation $U(\eta)=(\nu^2-1/4)/\eta^2$), we find that for values
of $C_2<2-0.1$ the model is disfavored  from observational data,
since the spectral index $n_s>1$, and for values of $C_2>2+0.1$
the ratio $r>0.2$, also the model is disfavored  from
observational data. In this form, for   deviations of the
value of $C_2=2$ (de Sitter inflation) our model does not work.

%Also, we note that the analytical solutions in terms of the Wittaker 
%functions and therefore for the power spectrums we have; 
%the constant $C_0$ becomes fundamental 
%in the solution of the variable $z$, and the solutions  for the modes $u_k$ and $v_k$, under the approximation
% $\dot{\varphi}\sim\dot{\varphi}_0$, we need $U(\eta)$ for the solution of $u_k$, and for the
  %analytical solution of $v_k$ is essential  the definition  of the variable $z$ for 
 %the tachyon field  Eq.(\ref{zt}), since $z\sim a$ (this does not occur in the standard field, 
 %since $z\sim a/H $ see Eq.(\ref{defz})). 

\section{Conclusions \label{conclu}}

In this paper we have analyzed an approach to exact solutions of
 cosmological perturbations
in the context of the Born-Infeld theory for two models; the
Easther's model, and the ansatz $U(\eta)$ given by
Eq.(\ref{potenc}). For both models, we have found explicit
solutions for the corresponding scalar field, effective tachyonic
potential, power spectrum of the curvature perturbations $P_{R}$,
scalar spectrum index $n_s$, power spectrum of the tensor
perturbations $\mathcal{P}_g$  and the tensor to scalar ratio $r$.

For the Easther's model in which $U=0$, we have obtained that only  the quantities
from the background field equations can be found analytically. In particular, we 
have got an effective tachyonic potential with the properties  $V_{,\varphi}(\varphi>0)<0$
and $V(\varphi\rightarrow \infty)\rightarrow 0$.
However, we have found that numerically the spectral index $n_s\simeq 3$, but  this 
value is disfavored from observational data. This result  for the spectral index is 
 of according with the numerical solution of $n_s=n_s(\varphi)$, if we assumed  that  the tachyon field slowly changes with the cosmic time,  
 and then 
  $n_s \equiv 3$.

For the model in which the  effective potential of cosmological perturbations is 
given by 
$U(\eta)=C_0+C_1/\eta+C_2/\eta^2$, we have found analytical results 
assuming  that the tachyon field $\varphi$ slowly changes with the time. Under 
this approximation we have obtained that the scalar and tensor perturbations can be written in terms of the 
Wittaker functions and essentially from definitions of $U(\eta)$ and $z$ for the tachyonic field, 
see Eq.(\ref{zt}).
  Considering the asymptotic limits for each one of the Fourier modes (scalar and tensor modes)
we have obtained the power spectrums scalar and tensorial in terms of the  modulus of the wavenumber $k$.
From these results we have considered the constraints on the parameters in our 
model from the Planck 2015 data. Here we have taken the constraint $r=r(n_s)$ plane or consistency relation, 
 and we have observed that for deviations of the value $C_2=2$ (de Sitter inflation)  our model is disfavored from 
observational data, and then  the model does not work. Also, we have noted from 
 the trajectories  $n_s-r$ plane, that  our model is well supported by the Planck data
only if the effective speed of sound  $c_{S_0}> 0.993$ (see Fig.\ref{figg5}) that is very similar to 
the speed of sound of 
the standard 
scalar field ($c_S\equiv 1$).  Therefore, our  ansatz of the effective potential of cosmological perturbations,  the tachyonic 
model works  for values of 
$c_{S_0}\simeq 1$ and $C_2\simeq $ 2.

\begin{acknowledgments}
R.H. was supported by Comisi\'on Nacional de Ciencias y
Tecnolog\'ia of Chile through FONDECYT Grant N$^{0}$ 1130628 and
DI-PUCV N$^{0}$ 123.724.
 R.G.P was supported by Proyecto Beca-Doctoral CONICYT N$^{0}$ 21100145

\end{acknowledgments}

%\\\\\\\\\\\\\\\\\\\\\\\\\\\\\\\\\\\\\\\\\\\\\\\\\\\\\\\\\\\\\\\\\\\\\\\

\end{document}